\documentclass[12pt]{article}
\usepackage{graphicx}
\usepackage{amsmath,amsthm,amsfonts,amssymb}
\usepackage{color}

\newcommand{\ba}{\begin{array}}
\newcommand{\ea}{\end{array}}
\newcommand{\nn}{\nonumber\\}

\newcommand{\C}{{\bf C}}

\newcommand{\del}{\partial}

\newcommand{\vvert}{\Big{\vert}}
\newcommand{\rar}{\rightarrow}

\newcommand{\fr}{\frac}

\newcommand{\scr}{\scriptsize}

\newcommand{\zb}{\bar{z}}

\makeatletter
\@addtoreset{equation}{section}

\makeatother

\begin{document}

\begin{titlepage}
\null
\begin{flushright}
\end{flushright}

\vskip 1.5cm
\begin{center}

 {\Large \bf  Soliton Scattering in Noncommutative Spaces}

\vskip 1.7cm
\normalsize

 {\large Masashi Hamanaka\footnote{E-mail:
 hamanaka@math.nagoya-u.ac.jp}
 and 
Hisataka Okabe\footnote{has worked at a company since April 2018}
 }

\vskip 1.5cm

        {\it Graduate School of Mathematics, Nagoya University,\\
                     Chikusa-ku, Nagoya, 464-8602, JAPAN}
\vskip 0.5cm

\vskip 1.5cm

{\bf \large Abstract}

\vskip 0.5cm
 
\end{center}
We discuss exact multi-soliton solutions to
integrable hierarchies on noncommutative space-times
in diverse dimension. 
The solutions are represented by quasi-determinants
in compact forms.
We study soliton scattering processes in 
the asymptotic region 
where the configurations could be real-valued.
We find that the asymptotic configurations 
in the soliton scatterings
can be all the same as commutative ones,
that is, the configuration of $N$-soliton solution has 
$N$ isolated localized lump of energy
and each solitary wave-packet lump preserves 
its shape and velocity in the scattering process.
The phase shifts are also the same as commutative ones.
As new results, we present multi-soliton solutions to 
noncommutative anti-self-dual Yang-Mills hierarchy
and discuss 2-soliton scattering in detail.
 

\end{titlepage}

\clearpage
\baselineskip 6mm

\section{Introduction}

Noncommutative  (NC) extension of integrable systems 
has attracted many researchers in both mathematics and physics
for long time. 
It would be a recent breakthrough that  
exact multi-soliton solutions to the noncommutative KP hierarchy
are constructed in terms of quasi-determinants \cite{EGR}. 
The quasideterminants are first introduced in 1991
by Gelfand and Retakh \cite{GeRe}
in the context of noncommutative generalization
of theory of determinants of matrices.
It has been found that the quasideterminants
play important roles in construction of
exact solutions to noncommutative integrable systems.
(See e.g. \cite{DoLe, EGR2, GHN, GHN2, GiMa, GiNi, GNO07, GNS, HaHa, 
Hamanaka_JHEP, Hamanaka_PS, LNT, ReRu, HSS} and references therein.)
It is interesting that quasideterminants
simplify proofs 
in commutative theories.

For the last several years, 
extension of integrable systems 
to noncommutative space-times
has been studied intensively.
This can be realized by using the star-product.  
{}From now on,, the word ``noncommutative'' is
assumed to refer to generalization to noncommutative spaces,
not to non-abelian and so on.
For surveys on integrable systems 
in noncommutative space-times, see e.g. \cite{
DiMH_proc,
Hamanaka_PhD,
Hamanaka_proc, HaTo3, 
Lechtenfeld_proc, Mazzanti, Tamassia}.

$N$-soliton solutions are stable 
in the sense that the configuration
has $N$ localized lump of energy
where shape and velocity of each localized lump keep intact
in the scattering process. 
Existence of them closely relates to
existence of infinite conserved quantities or
infinite dimensional symmetry. 
Hence it is worth studying the stability
of the noncommutative soliton dynamics. 
In the star-product formalism,
space-time coordinates and
functions take c-number values and 
scattering dynamics can be clarified
explicitly.
However, there are few studies on it
\cite{DiMH_KdV, MaPa, Paniak}.

In this paper, we study exact multi-soliton solutions to
noncommutative integrable hierarchies and 
the asymptotic behavior of them 
where the asymptotic configurations are real-valued.
We focus on the noncommutative Korteweg-de Vries (KdV), 
Kadomtsev-Petviashvili (KP) and anti-self-dual Yang-Mills (ASDYM)
equations in $(1+1)$, $(2+1)$ and 4 dimension, respectively.
We find that 
the asymptotic behavior in soliton scatterings
is all the same as commutative ones,
that is, the $N$-soliton solution
has $N$ isolated localized lump of energy 
and each wave-packet lump  preserves its shape and velocity
in the scattering process.
The phase shift is also the same as commutative one.
The analysis of 2-soliton scattering in 
the noncommutative ASDYM equation is new.
Property of the quasideterminants and the star products
plays crucial roles. 

This paper is organized as follows.
In section 2, we give a brief introduction to
noncommutative field theory in the star-product formalism.
In section 3, we make a brief review 
of the quasi-determinants.
In section 4, we define
noncommutative integrable hierarchy and
construct exact multi-soliton solutions 
by using quasideterminants. 
Asymptotic behaviors of noncommutative
KdV and KP solitons are discussed 
in the star-product formalism \cite{Hamanaka_JHEP}. 
In section 5, we define noncommutative ASDYM hierarchy
and give exact solutions to it in terms of quasideterminants. 
Asymptotic behavior of 2-soliton solutions
are discussed. This section gives new results. 

\section{Integrable Equations in Noncommutative Spaces}

Noncommutative spaces are defined
by the noncommutativity of the coordinates:
\begin{eqnarray}
\label{nc_coord}
[x^\mu,x^\nu]=i\theta^{\mu\nu},
\end{eqnarray}
where the constant $\theta^{\mu\nu}$ is
called the {noncommutative parameter}.
If the coordinates are real,
noncommutative parameters should be real
because of hermicity of the coordinates.
We note that the noncommutative parameter $\theta^{\mu\nu}$ is
anti-symmetric with respect to $\mu$ and $\nu$ 
which implies that the rank of it is even.
In $(1+1)$-dimension with the coordinate $(t,x)$,
there is unique choice of noncommutativity : $[x,t]=i\theta$
which is space-time noncommutativity.
In $(2+1)$-dimension with the coordinate $(t,x,y)$,
there are essentially two kind of
choices of noncommutativity, that is,
space-space noncommutativity: $[x,y]=i\theta$
and space-time noncommutativity: $[x,t]=i\theta$
or $[y, t]=i\theta$.

Noncommutative field theories are given by the replacement of ordinary products
in the commutative field theories with the {star-products}.
The star-product is defined for ordinary fields.
On flat spaces, is it represented explicitly by
\begin{eqnarray}
f\star g(x)&:=&\mbox{exp}
\left(\frac{i}{2}\theta^{\mu\nu} \partial^{(x_1)}_\mu
\partial^{(x_2)}_\nu \right)
f(x_1)g(x_2)\Big{\vert}_{x_1=x_2=x}\nonumber\\
&=&f(x)g(x)+\frac{i}{2}\theta^{\mu\nu}\partial_\mu f(x)\partial_\nu g(x)
+{\cal{O}} (\theta^2),
\label{star}
\end{eqnarray}
where $\del_\mu^{(x)}:=\del/\del x^{\mu}$.
This is known as the {Moyal product} \cite{Moyal}.
The ordering of fields in nonlinear terms are determined
so that some structures such as gauge symmetries 
should be preserved.

The star-product has associativity:
$f\star(g\star h)=(f\star g)\star h$. 
It reduces to the ordinary product 
in the commutative limit:  $\theta^{\mu\nu}\rar 0$.
In this sense, the noncommutative field theories
are deformed theories from the commutative ones.
The replacement of the product  makes the ordinary
spatial coordinates ``noncommutative,''
that is,
$[x^\mu,x^\nu]_\star:=x^\mu\star x^\nu-x^\nu\star x^\mu=i\theta^{\mu\nu}$.

We note that the fields themselves take c-number values
and the differentiation and the integration for them
are the same as commutative ones. 

Here is a gallery of noncommutative integrable equations.
Time and spatial coordinates are
denoted by $t$ and $x,y$, respectively.
\begin{itemize}

 \item In $(1+1)$ dimension:
\begin{itemize}
 \item Noncommutative KdV equation
\begin{eqnarray}
\label{ncKdV}
 \dot{u}=\frac{1}{4}u^{\prime\prime\prime}+\frac{3}{4}
\left(u^\prime \star u + u \star u^\prime \right),
\end{eqnarray}
where   $\dot{u}:=\del f/\del t,~u^\prime:=\del f/\del x,~
u^{\prime\prime}:=\del^2 f/\del x^2$ and so on. 
2-soliton dynamics is discussed \cite{DiMH_KdV}.

 \item Noncommutative Boussinesq equation 
\begin{eqnarray}
 3\ddot{u}+u^{\prime\prime\prime\prime}+2
  (u\star u)^{\prime\prime}-2[u,\del_x^{-1}\dot{u}]_\star^\prime=0.
\end{eqnarray}

 \item Noncommutative Non-Linear Schr\"odinger equation
\begin{eqnarray}
 \label{nls}
i\dot{\psi}=\psi^{\prime\prime}-2\varepsilon\psi \star \bar{\psi} \star \psi.
\end{eqnarray}

 \item Noncommutative modified KdV equation
\begin{eqnarray}
 \dot{v}=\frac{1}{4}v^{\prime\prime\prime}
  -\frac{3}{4}\left(v\star v\star v^{\prime}+
v^{\prime}\star v\star v \right).
\end{eqnarray}
This is connected with the noncommutative KdV equation
via the noncommutative Miura map: $u=v^\prime -v^2$ \cite{DiMH_KdV}.

 \item Noncommutative Burgers equation 
\begin{eqnarray}
 \dot{u}=u^{\prime\prime}+2u\star u^{\prime}~~~{\mbox{ or }}~~~
 \dot{u}=u^{\prime\prime}-2 u^{\prime}\star u.
\end{eqnarray}
These can be linearized by the noncommutative Cole-Hopf transformations:
$u=\psi^{-1} \star\psi^\prime $ or $u=-\psi^\prime\star \psi^{-1}$,
respectively. (See e.g. \cite{HaTo2}.) 
2 shock-wave dynamics is discussed \cite{MaPa}.       

\end{itemize}

 \item In $(2+1)$ dimension: 
\begin{itemize}

 \item Noncommutative KP equation 
\begin{eqnarray}
\label{ncKP}
 u_t=\frac{1}{4}u_{xxx}+\frac{3}{4}
 \left(u_x\star u + u \star u_x \right)
 +\partial_x^{-1} u_{yy}-[u, \partial_x^{-1}u_y]_\star.
\end{eqnarray}
where the subscripts denote partial derivatives and 
$\partial_x^{-1}f(x)=\int^x dx^\prime f(x^\prime)$.
This reduces to 
the KdV equation in $(1+1)$-dimension by $\partial_y=0$  .
2-soliton dynamics is discussed in detail \cite{Paniak}.
       
 \item Noncommutative Zakharov system \cite{Hamanaka_NPB}
\begin{eqnarray*}
 i\psi_t&=&\psi_{xy}-\varepsilon \psi \star
  \del_x^{-1} \del_{y}(\bar{\psi} \star \psi)
  -\varepsilon \del_x^{-1}\del_{y}(\psi\star \bar{\psi})\star \psi,
\end{eqnarray*}
where $\varepsilon=\pm 1$.
This reduces to 
the noncommutative Non-Linear Schr\"odinger (NLS)  equation
       in $(1+1)$-dimension by $x=y$  .

 \item Noncommutative Bogoyavlenskii-Calogero-Schiff equation \cite{Toda}
\begin{eqnarray*}
4 u_t=\displaystyle u_{xxy}
+2\left(u\star u\right)_{y}
+ \left(u_{y}\star  \del_x^{-1}u+\del_x^{-1}u\star u_{y}\right)
+\del_x^{-1}[u,\del_x^{-1}[u, \del_x^{-1} u_{y}]_\star]_\star.
\label{bcs} 
\end{eqnarray*}
This reduces to the noncommutative KdV equation by $x=y$  .

 \item Noncommutative Davey-Stewartson equation \cite{Hamanaka_PLB}
\begin{eqnarray*}
\left\{
\begin{array}{l}
2i q_t=(\del_x^2-\del_y^2)q
+R_1\star q-q\star R_2,\\
2i r_t=-(\del_x^2-\del_y^2)r
+R_2\star r-r\star R_1,
\end{array}\right.
\end{eqnarray*}
where  $(\del_x-i\del_y)R_1=-(\del_x+i\del_y)(q\star r),~
 (\del_x+i\del_y)R_2=(\del_x-i\del_y)(r\star q)$.
This reduces to 
the noncommutative NLS equation in $(1+1)$-dimension by $\partial_y=0$
and $R_1=-q\star r, R_s=r\star q, q=\psi, r=\bar{\psi}$.
      
\end{itemize}

 \item  In 4-dimension, there is an important integrable equation:
	noncommutative anti-self-dual Yang-Mills (ASDYM) equation
\begin{eqnarray}
 \label{ncASDYM}
 F^\star_{zw}=0,~F^\star_{\bar{z}\bar{w}}=0,~
 F^\star_{z\bar{z}} +  F^\star_{w\bar{w}}=0,
\end{eqnarray}
where $z$ and $w$ denote
local coordinates of the 4-dimensional Euclidean 
plane $\mathbb{C}^2$, and 
$F^\star_{\mu\nu}:=\partial_\mu A_\nu-\partial_\nu A_\mu
+[A_\mu,A_\nu]_\star$ denotes the field strength. 
There are two choices of rank 2 and 4 with respect to noncommutativity.
We note that the noncommutative ASDYM equation gives rise by reduction
to various noncommutative lower-dimensional integrable equations
including all equations above except for noncommutative Burgers equation.
More examples are summarized 
in \cite{Hamanaka_PLB, Hamanaka_NPB}, 
which would be 
evidence for noncommutative version \cite{HaTo}
of the Ward conjecture \cite{Ward}. (See also \cite{AbCl, MaWo}.)

\end{itemize}

\section{Review of Quasi-determinants}

In this section, we briefly
review quasi-determinants introduced by Gelfand and Retakh
\cite{GeRe} and present a few properties
of them which play important roles in the following sections.
The detailed discussion is seen in e.g. \cite{GGRW}.

Quasi-determinants are not just a generalization of
usual commutative determinants but rather
related to inverse matrices. From now on,
we assume existence of the inverses in any case.

Let $A=(a_{ij})$ be a $N\times N$ matrix and 
$B=(b_{ij})$ be the inverse matrix of $A$,
that is, $A\star B=B\star A =1$.
In this paper, all products of matrix elements are assumed to be
star-products. 

Quasi-determinants of $A$ are defined formally
as the inverse of the elements of $B=A^{-1}$:
\begin{eqnarray}
 \vert A \vert_{ij}:=b_{ji}^{-1}.
\end{eqnarray}
In the commutative limit, this reduces to
\begin{eqnarray}
 \vert A \vert_{ij} \stackrel{\theta\rightarrow 0}{\longrightarrow}
  (-1)^{i+j}\frac{\det A}{\det {A}^{ij}},
\label{limit}
\end{eqnarray}
where ${A}^{ij}$ is the matrix obtained from $A$
deleting the $i$-th row and the $j$-th column.

We can write down more explicit form of quasi-determinants.
In order to see it, let us recall the following formula
for the inverse $2\times 2$ block matrix:
\begin{eqnarray*}
 \left[
 \begin{array}{cc}
  A&B \\C&d
 \end{array}
 \right]^{-1}
=\left[\begin{array}{cc}
A^{-1}+A^{-1}\star B\star S^{-1}\star C \star A^{-1}
 &-A^{-1}\star B\star S^{-1}\\
 -S^{-1}\star C\star A^{-1}
&S^{-1}
\end{array}\right],
\end{eqnarray*}
where $A$ is a square matrix and $d$ is a single element and 
$S:=d-C\star A^{-1}\star B$ is called the Schur complement. 
We note that any matrix can be decomposed
as a $2\times 2$ matrix by block decomposition
where one of the diagonal parts is $1 \times 1$.
We note that 
by choosing an appropriate partitioning,
any element in the inverse of a square matrix can be expressed
as the inverse of the Schur complement.
Hence quasi-determinants can be defined iteratively by:
\begin{eqnarray}
 \vert A \vert_{ij}&=&a_{ij}-\sum_{i^\prime (\neq i), j^\prime (\neq j)}
  a_{ii^\prime} \star (({A}^{ij})^{-1})_{i^\prime j^\prime} \star
  a_{j^\prime
  j}\nonumber\\
 &=&a_{ij}-\sum_{i^\prime (\neq i), j^\prime (\neq j)}
  a_{ii^\prime} \star (\vert {A}^{ij}\vert_{j^\prime i^\prime })^{-1}
  \star a_{j^\prime j}.
\end{eqnarray}
It is convenient to represent the quasi-determinant
as follows:
\begin{eqnarray}
 \vert A\vert_{ij}=
  \begin{array}{|ccccc|}
   a_{11}&\cdots &a_{1j} & \cdots& a_{1n}\\
   \vdots & & \vdots & & \vdots\\
   a_{i1}&~ & {\fbox{$a_{ij}$}}& ~& a_{in}\\
   \vdots & & \vdots & & \vdots\\
   a_{n1}& \cdots & a_{nj}&\cdots & a_{nn}
  \end{array}~.
\end{eqnarray}

Examples of quasi-determinants are,
for a $1\times 1$ matrix $A=a$
 \begin{eqnarray*}
  \vert A \vert  = a,
 \end{eqnarray*}
and 
for a $2\times 2$ matrix $A=(a_{ij})$
 \begin{eqnarray*}
  \vert A \vert_{11}=
   \begin{array}{|cc|}
   \fbox{$a_{11}$} &a_{12} \\a_{21}&a_{22}
   \end{array}
 =a_{11}-a_{12}\star a_{22}^{-1}\star a_{21},~~~
  \vert A \vert_{12}=
   \begin{array}{|cc|}
   a_{11} & \fbox{$a_{12}$} \\a_{21}&a_{22}
   \end{array}
 =a_{12}-a_{11}\star a_{21}^{-1}\star a_{22},\nonumber\\
  \vert A \vert_{21}=
   \begin{array}{|cc|}
   a_{11} &a_{12} \\ \fbox{$a_{21}$}&a_{22}
   \end{array}
 =a_{21}-a_{22}\star a_{12}^{-1}\star a_{11},~~~
  \vert A \vert_{22}=
   \begin{array}{|cc|}
   a_{11} & a_{12} \\a_{21}&\fbox{$a_{22}$}
   \end{array}
 =a_{22}-a_{21}\star a_{11}^{-1}\star a_{12}, 
 \end{eqnarray*}
 and for a $3\times 3$ matrix $A=(a_{ij})$
  \begin{eqnarray*}
  \vert A \vert_{11}
   &=&
   \begin{array}{|ccc|}
   \fbox{$a_{11}$} &a_{12} &a_{13}\\ a_{21}&a_{22}&a_{23}\\a_{31}&a_{32}&a_{33}
   \end{array}
=a_{11}-(a_{12}, a_{13})\star \left(
\begin{array}{cc}a_{22} & a_{23} \\a_{32}&a_{33}\end{array}\right)^{-1}
\star \left(
\begin{array}{c}a_{21} \\a_{31}\end{array}
\right)
\nonumber\\
  &=&a_{11}-a_{12}\star  \begin{array}{|cc|}
                   \fbox{$a_{22}$} & a_{23} \\a_{32}&a_{33}
                   \end{array}^{-1}  \star a_{21}
           -a_{12}\star \begin{array}{|cc|}
                   a_{22} & a_{23} \\\fbox{$a_{32}$}&a_{33}
                   \end{array}^{-1} \star a_{31}      \nonumber\\
&&~~~~    -a_{13}\star \begin{array}{|cc|}
                   a_{22} & \fbox{$a_{23}$} \\a_{32}&a_{33}
                                         \end{array}^{-1}\star  a_{21}
           -a_{13}\star \begin{array}{|cc|}
                   a_{22} & a_{23} \\a_{32}&\fbox{$a_{33}$}
                   \end{array}^{-1} \star a_{31},
 \end{eqnarray*}
and so on.

Quasideterminants have various interesting properties
similar to those of determinants.
The following ones are relevant to 
the discussion on soliton scattering.

\vspace{3mm}
\noindent
{\bf Proposition 3.1 \cite{GeRe}}
Let $A=(a_{ij})$ be a square matrix of order $n$.

\noindent 
(i) {Permutation of Rows and Columns}.

The quasi-determinant $\vert A\vert_{ij}$
does not depend on permutations of rows and columns
in the matrix $A$ that do not involve the $i$-th row
and $j$-th column.

\noindent
(ii) {The multiplication of rows and columns}.

Let the matrix $M=(m_{ij})$ be obtained from
the matrix $A$ by multiplying the $i$-th row by
$f(x)$ from the left, that is,
$m_{ij}=f \star a_{ij}$ and $m_{kj}=a_{kj}$
for $k\neq i$. Then
\begin{eqnarray}
 \vert M\vert_{kj}=\left\{
                   \begin{array}{ll}
                    f \star\vert A \vert_{ij}& \mbox{for}~ k=i 
                     \\ \vert A \vert_{kj}& \mbox{for}~ k\neq i~~\mbox{and
                     }~f~ \mbox{is invertible}
                   \end{array}\right.
\end{eqnarray}

Let the matrix $N=(n_{ij})$ be obtained from
the matrix $A$ by multiplying the $j$-th column by
$f(x)$ from the right, that is,
$n_{ij}=a_{ij}\star f $ and $n_{il}=a_{il}$
for $l\neq j$. Then
\begin{eqnarray}
 \vert N\vert_{il}=\left\{
                   \begin{array}{ll}
                    \vert A \vert_{ij}\star f& \mbox{for}~ l=j
                     \\ \vert A \vert_{il}& \mbox{for}~ l\neq j~~\mbox{and
                     }~ f ~\mbox{is invertible} 
                   \end{array}\right.
\end{eqnarray}

\vspace{3mm}

\section{NC Integrable Hierarchy and Soliton Solutions}

In this section, we give exact multi-soliton
solutions to noncommutative integrable hierarchies
in terms of quasi-determinants.
In the commutative case, determinants of
Wronski matrices play crucial roles.
In the noncommutative case, 
quasi-determinants give a better formulation. 
We review foundation of the noncommutative KP hierarchy and
the reduced hierarchies,
so called noncommutative Gelfand-Dickey (GD) hierarchies, 
and present the exact multi-soliton solutions to them
developed by Etingof, Gelfand and Retakh \cite{EGR}. (See also \cite{GiNi})

\vspace{3mm}

An $N$-th order pseudo-differential operator $A$
is represented as follows
\begin{eqnarray}
 A=a_N \del_x^N + a_{N-1}\del_x^{N-1}+ \cdots
+ a_0 +a_{-1}\del_x^{-1}+a_{-2}\del_x^{-2}+\cdots,
\end{eqnarray}
where $a_i$ is a function of $x$ 
associated with noncommutative associative products
(here, the star products).
When the coefficient of the highest order $a_N$ equals to 1,
we call it monic.
Here we introduce the following symbols:
\begin{eqnarray}
 A_{\geq r}&:=& \del_x^N + a_{N-1}\del_x^{N-1}+ \cdots + a_{r}\del_x^{r},\\
 A_{\leq r}&:=& A - A_{\geq r+1}
 = a_{r}\del_x^{r} + a_{r-1}\del_x^{r-1} +\cdots.
\end{eqnarray}

The action of a differential operator $\partial_x^n$ on
a multiplicity operator $f$ is formally defined
as the following generalized Leibniz rule:
\begin{eqnarray}
 \partial_x^{n}\cdot f:=\sum_{i\geq 0}
\left(\begin{array}{c}n\\i\end{array}\right)
(\partial_x^i f)\partial^{n-i},
\end{eqnarray}
where the binomial coefficient is given by
\begin{eqnarray}
\label{binomial}
 \left(\begin{array}{c}n\\i\end{array}\right):=
\frac{n(n-1)\cdots (n-i+1)}{i(i-1)\cdots 1}.
\end{eqnarray}
We note that the definition of the binomial coefficient (\ref{binomial})
is applicable to the case of negative $n$, which
implies that the action of
negative power of differential operators is defined.

The composition of pseudo-differential operators
is also well-defined and the total set
of pseudo-differential operators forms
an operator algebra.
For a monic pseudo-differential operator $A$, there exist
the unique inverse $A^{-1}$ and the unique $m$-th root $A^{1/m}$
which commute with $A$.
(These proofs are all the same as commutative ones as far as 
the commutative limit exists.)
For more on pseudo-differential operators
and Sato's theory, see e.g. \cite{BBT, Blaszak, Dickey, Kupershmidt}.

\vspace{3mm}

\subsection{Noncommutative KP and KdV hierarchies}

In order to define the noncommutative KP hierarchy,
let us introduce a monic pseudo-differential operator:
\begin{eqnarray}
 L = \partial_x + u_2 \partial_x^{-1}
 + u_3 \partial_x^{-2} + u_4 \partial_x^{-3} + \cdots,
\end{eqnarray}
where the coefficients $u_k$ ($k=2,3,\ldots$) are functions
of infinite coordinates $\vec{x}:=(x_1,x_2,\ldots)$ with $x_1\equiv x$:
\begin{eqnarray}
 u_k=u_k(x_1,x_2,\ldots).
\end{eqnarray}
The noncommutativity is introduced into
the coordinates $(x_1,x_2,\ldots)$ as Eq. (\ref{nc_coord}) here.

In order to define the noncommutative KP hierarchy,
let us introduce a differential operator $B_m$ as follows:
\begin{eqnarray}
 B_m:=(\underbrace{L\star \cdots \star L}_{ m{\scriptsize\mbox{
     times}}})_{\geq 0}=:(L^m)_{\geq 0}.
\end{eqnarray}
The noncommutative KP hierarchy is defined as follows:
\begin{eqnarray}
 \del_m L = \left[B_m, L\right]_\star,~~~m=1,2,\ldots,
\label{lax_sato}
\end{eqnarray}
where the action of $\del_m:=\del/\del x_m$
on the pseudo-differential operator $L$
is defined by $\del_m L :=[\del_m,L]_\star$ or $\del_m \del_x^k=0$.
The KP hierarchy gives rise to a set of infinite differential
equations with respect to infinite kind of fields from the
coefficients in Eq. (\ref{lax_sato}) for a fixed $m$. Hence it
contains huge amount of differential (evolution) equations for all
$m$. The LHS of Eq. (\ref{lax_sato}) becomes $\del_m u_k$ which
shows a kind of flow in the $x_m$ direction.

If we put the constraint $(L^l)_{\leq -1}=0$ or equivalently
$L^l=B_l~(l=2,3,\cdots)$ 
on the noncommutative KP hierarchy (\ref{lax_sato}), 
we get a reduced noncommutative KP hierarchy which is called 
the {l-reduction} of the noncommutative KP hierarchy, or
the {noncommutative $l$KdV hierarchy}, or the $l$-th 
{noncommutative GD hierarchy}. 
In particular, the 2-reduction of noncommutative KP hierarchy
is just the noncommutative KdV hierarchy.
We can easily show
\begin{eqnarray}
\label{Nl}
\frac{\partial u_k}{\partial x_{nl}}=0,
\end{eqnarray}
for all $n,k$ because $
\partial L^l/\partial x_{nl}=[B_{nl},L^l]=[(L^{l})^n,L^l]=0$. 
This time, the constraint $L^l=B_l$ gives simple relationships which make it
possible to represent infinite kind of fields
$u_{l+1},u_{l+2},u_{l+3},\ldots$ in terms of $(l-1)$ kind of
fields $u_{2},u_{3},\ldots, u_{l}$. (cf. Appendix A in
\cite{Hamanaka_JMP}.) 

{}Let us see explicit examples.
\begin{itemize}

\item Noncommutative KP hierarchy

The coefficients of each powers of (pseudo-)differential operators
in the noncommutative KP hierarchy (\ref{lax_sato}) yield a series of infinite
noncommutative ``evolution equations,'' that is, for $m=1$
\begin{eqnarray}
\partial_x^{1-k})~~~ \del _1 u_{k}=u_{k}^\prime,~~~k=2,3,\ldots
~~~\Rightarrow~~~x^1\equiv x,
\end{eqnarray}
for $m=2$
\begin{eqnarray}
\label{KP_hie}
\partial_x^{-1})~~~\del_2 u_{2}
&=&u_2^{\prime\prime}+2u_{3}^{\prime},\nonumber \\
\partial_x^{-2})~~~
\del_2 u_{3}&=&u_3^{\prime\prime}+2u_4^{\prime}
+2u_2\star u_2^\prime +2[u_2,u_3]_\star,\nonumber \\
\partial_x^{-3})~~~
\del_2 u_{4}&=&u_{4}^{\prime\prime}+2u_{5}^{\prime}
+4u_3\star u_2^\prime-2u_2\star u_2^{\prime\prime}
+2[u_2,u_4]_\star,\nn
\partial_x^{-4})~~~\del_2 u_{5}&=&\cdots,
\end{eqnarray}
and for $m=3$
\begin{eqnarray}
\label{3flow}
\partial_x^{-1})~~~
\del_3 u_{2}&=&u_{2}^{\prime\prime\prime}+3u_3^{\prime\prime}
+3u_4^{\prime}+3u_2^\prime\star u_2+3u_2\star u_2^\prime,
\nonumber\\
\partial_x^{-2})~~~
\del_3 u_{3}&=&u_{3}^{\prime\prime\prime}+3u_{4}^{\prime\prime}
+3u_{5}^\prime+6u_{2}\star u_{3}^\prime+3u_2^\prime\star u_3
+3u_3\star u_2^\prime+3[u_2, u_4]_\star,\nn
\partial_x^{-3})~~~
\del_3 u_{4}&=&u_{4}^{\prime\prime\prime}+3u_{5}^{\prime\prime}
+3u_{6}^\prime+3u_{2}^\prime \star u_{4}+3u_2\star u_4^\prime
+6u_4\star u_2^\prime\nn
&&-3u_2\star u_3^{\prime\prime}
-3u_3\star u_2^{\prime\prime}+6u_3\star u_3^{\prime}
+3[u_2,u_5]_\star+3[u_3,u_4]_\star,\nn
\partial_x^{-4})~~~\del_3 u_{5}&=&\cdots.
\end{eqnarray}
These contain the $(2+1)$-dimensional noncommutative KP equation
\eqref{ncKP} with $2u_2\equiv u, x_2\equiv
y,x_3\equiv t$ and $\partial_x^{-1}f(x)=\int^x dx^\prime f(x^\prime)$.
We note that infinite kind of fields $u_3, u_4, u_5,\ldots$
are represented in terms of one kind of field  $2u_2\equiv u$
as is seen in Eq. (\ref{KP_hie}).

\item Noncommutative KdV Hierarchy (2-reduction of the noncommutative KP hierarchy)

Putting the constraint $L^2=B_2=:\del_x^2+u$ on 
the noncommutative KP hierarchy, we get the noncommutative KdV hierarchy.
We note that the even-th flows are trivial. 
The following noncommutative hierarchy
\begin{eqnarray}
\label{KdV_hie}
 \frac{\partial u}{\partial x^m}=\left[B_m, L^2\right]_\star,
\end{eqnarray}
has neither positive nor negative power of
(pseudo-)differential operators for the same reason as commutative
case and gives rise to the $m$-th KdV equation for each
$m=1,3,5,\cdots$.
The noncommutative KdV hierarchy (\ref{KdV_hie})
coincides with the $(1+1)$-dimensional noncommutative KdV equation
for $m=3$ with $x_3\equiv t$,
\eqref{ncKdV} and
and with the $(1+1)$-dimensional 5-th noncommutative KdV equation \cite{Toda}
for $m=5$ with $x_5\equiv t$:
\begin{eqnarray}
\dot{u}&=&\frac{1}{16}u^{\prime\prime\prime\prime\prime}
+\frac{5}{16}(u\star
u^{\prime\prime\prime}+u^{\prime\prime\prime}\star u)
+\frac{5}{8}(u^{\prime}\star u^{\prime}+u\star u\star u)^\prime.
\end{eqnarray}
\end{itemize}
In this way, we can generate infinite set of the $l$-reduced noncommutative KP
hierarchies. 
Explicit examples are seen in e.g. \cite{Hamanaka_JMP}.
(See also \cite{CSS, KoOe, Ma, OlSo, Wang, WaWa}.)

\subsection{Multi-soliton Solutions to NC KP and KdV hierarchies}

Now we construct multi-soliton solutions
of the noncommutative KP hierarchy. 
Let us introduce the following functions,
\begin{eqnarray}
 \label{argument}
 f_s(\vec{x})=e_\star^{\xi(\vec{x};k_s)}
  +a_s e_\star^{\xi(\vec{x};k^{\prime}_s)},~~~
\mbox{where}~~~
 \xi(\vec{x};k)=x_1k+x_2 k^2+x_3 k^3+\cdots,
\end{eqnarray}
where $k_s$, $k^{\prime}_s$ and $a_s$ are constants. 
Star exponential functions are defined by
\begin{eqnarray}
 e_\star^{f(x)}:=1+\sum_{n=1}^{\infty}\frac{1}{n!}
\underbrace{f(x)\star \cdots \star f(x)}_{n ~{\mbox{\scriptsize times}}}.
\end{eqnarray}

An $N$-soliton solution to the noncommutative KP hierarchy (\ref{lax_sato})
is given by \cite{EGR},
\begin{eqnarray}
 L=\Phi_N \star \partial_x \Phi_N^{-1},
\label{Nsol1}
\end{eqnarray}
where
\begin{eqnarray}
 \Phi_N \star f&=&\vert W(f_1, \ldots, f_N, f)\vert_{N+1,N+1},\nonumber\\
&=&
\begin{array}{|ccccc|}
 f_1&f_2 & \cdots& f_N & f\\
 f^\prime_1&  f^\prime_2& \cdots& f^\prime_N&f^\prime\\
 \vdots& \vdots&\ddots & \vdots &\vdots\\
 f^{(N-1)}_1& f^{(N-1)}_2& \cdots & f^{(N-1)}_N &f^{(N-1)}\\
 f^{(N)}_1& f^{(N)}_2& \cdots &f^{(N)}_N & \fbox{$f^{(N)}$}\\
      \end{array}~ .
\label{sol_KP}
\end{eqnarray}
The Wronski matrix $W(f_1,f_2,\cdots, f_m)$ is given by
\begin{eqnarray}
 W(f_1,f_2,\cdots, f_m):=
\left[\begin{array}{cccc}
 f_1&f_2 & \cdots& f_m\\
 f^\prime_1& f^\prime_2& \cdots& f^\prime_m\\
 \vdots& \vdots&\ddots & \vdots\\
  f^{(m-1)}_1& f^{(m-1)}_2& \cdots & f^{(m-1)}_m\\
      \end{array}\right],
\end{eqnarray}
where $f_1,f_2, \cdots, f_m$ are functions of $x$
and $f^\prime:=\del f/\del x,~
f^{\prime\prime}:=\del^2 f/\del x^2,~
f^{(m)}:=\del^m f/\del x^m$ and so on. 

In the commutative limit, $\Phi_N \star f$ is reduced to
\begin{eqnarray}
 \Phi_N \star f \longrightarrow
  \frac{\det W(f_1,f_2,\ldots,f_N,f)}
{\det W(f_1,f_2,\ldots,f_N)},
\end{eqnarray}
which just coincides with the commutative $N$-soliton solution \cite{Dickey}. 
In this respect, quasi-determinants are fit
to this framework of the Wronskian solutions,
however, give a new formulation of it. 

{}From Eq. (\ref{Nsol1}), we have a more explicit form as 
\begin{eqnarray}
 u_2=\partial_x \left(\sum_{s=1}^{N} W^\prime_s \star W_s^{-1}  \right),
\label{Nsol2}
~~~\mbox{where}~~~
 W_s:=\vert W(f_1,\ldots,f_s)\vert_{ss}.
\end{eqnarray}
In the soliton solutions, 
The $l$-reduction condition $(L^l)_{\leq -1}=0$ or $L^l=B_l$
is equivalent to the constraint 
$k_s=\epsilon k^{\prime}_s~(s=1,2,\cdots,N)$,
where $\epsilon$ is the $l$-th root of unity.

\vspace{3mm}

\subsection{Asymptotic Behavior of the Exact Soliton Solutions}

In this subsection, we discuss asymptotic behavior of
the $N$-soliton solutions in asymptotic region of 
infinitely past and future.
In the star-product formalism,
all coordinates are regarded as c-number functions.
We can as usual plot the configurations and
interpret the positions of localized wave packet lump,
and read phase shifts.
Here we restrict ourselves to noncommutative KdV
and KP equations (the third flow of the hierarchies)
with space-time noncommutativity $[x,t]_\star=i\theta$
where $(x,t)\equiv (x_1, x_3)$. 
Discussion to other noncommutative hierarchies is similarly made 
\cite{Hamanaka_JHEP}.

First, let us comment on an important formula 
which is relevant to one-soliton solutions.
Let $x,t$ be noncommutative space-time coordinates.
Introducing new noncommutative coordinates as $z:=x+v t,
\zb:=x-v t$, we can easily find
\begin{eqnarray}
 f(z)\star g(z)= f(z) g(z)
\end{eqnarray}
because the star-product (\ref{star}) is rewritten in terms of
$(z,\zb)$ as 
\begin{eqnarray}
 f(z,\zb)\star g(z,\zb)=
e^{iv\theta\left(
\partial_{\zb_1}
\partial_{z_2}-
\partial_{z_1}
\partial_{\zb_2}
\right)}f(z_1,\zb_1)
g(z_2,\zb_2) \Big{\vert}_{\scr
\begin{array}{c} z_1=z_2=z\\
\zb_1 =\zb_2=\zb. \end{array}}
\label{hol}
\end{eqnarray}
Hence noncommutative one soliton-solutions can be
the same as commutative ones.

\vspace{3mm}

When $f(x)$ is a linear function,
the star exponential function $e_\star^{f(x)}$ is tractable
because it satisfies 
\begin{eqnarray}
\label{inverse_exp}
 (e_\star^{\xi(\vec{x};k)})^{-1} &=& e_\star^{-\xi(\vec{x};k)},\\
 \partial_x e_\star^{\xi(\vec{x};k)}
  &=& k e_\star^{\xi(\vec{x};k)}.
\label{der_exp}
\end{eqnarray}
These formula play crucial roles in discussion
on asymptotic behavior of the $N$-soliton solutions.

\subsubsection{Asymptotic behavior of noncommutative KdV solitons}

First, let us discuss 
the asymptotic behavior of the $N$-soliton solutions
to the noncommutative KdV equation.
The noncommutative KdV hierarchy is
the 2-reduction of the noncommutative KP hierarchy
and realized by putting $k^{\prime}_s=-k_s$ on the 
$N$-soliton solutions to the noncommutative KP hierarchy.
Here the constants $k_s$ and $a_s$ are non-zero real numbers
and $a_s$ is positive.
Because of the permutation property of the columns of quasi-determinants
in Proposition 3.1 (i),
we can assume $k_1< k_2 <\cdots <k_N$.

Let us discuss the soliton solutions to the noncommutative KdV
equation where the 
coordinates are specified as $(x,t)\equiv(x_1,x_3)$.
Let us define a new coordinate $X:=x+k_I^{2} t$
comoving with the $I$-th soliton and take $t\rightarrow \pm \infty$
limit. We note that $X$ is finite at any time.
Then, because of $x+k_s^{2} t=x+k_I^{2} t + (k_s^{2}
-k_I^{2})t$,
either $e_\star^{k_{s}(x + k_{s}^{2} t)}$
or $e_\star^{-k_{s}(x + k_{s}^{2} t)}$
goes to zero for $s\neq I$.
Hence the behavior of $f_s$ becomes at $t\rightarrow +\infty$:
\begin{eqnarray}
 f_s(\vec{x})\longrightarrow
\left\{\begin{array}{ll}
a_s e_\star^{- k_{s}(x + k_{s}^{2} t)} &s< I\\
e_\star^{k_{I}(x + k_{I}^{2} t)}
+a_I e_\star^{-k_{I}(x + k_{I}^{2} t)}&s=I\\
 e_\star^{ k_{s}(x + k_{s}^{2} t)} &s> I,\\
       \end{array}
\right. 
\end{eqnarray}
and at $t\rightarrow -\infty$:
\begin{eqnarray}
 f_s(\vec{x})\longrightarrow
\left\{\begin{array}{ll}
 e_\star^{ k_{s}(x + k_{s}^{2} t)} &s< I\\
e_\star^{k_{I}(x + k_{I}^{2} t)}
+a_I e_\star^{-k_{I}(x + k_{I}^{2} t)}&s=I\\
a_s e_\star^{- k_{s}(x + k_{s}^{2} t)} &s> I.\\
       \end{array}
\right. 
\end{eqnarray}
We note that the $s$-th ($s\neq I$) column
is proportional to a single 
exponential function
$e_\star^{\pm k_{s}(x + k_{s}^{2} t)}$ due to
Eq. (\ref{der_exp}).
Because of the multiplication property of
columns of quasi-determinants in Proposition 3.1 (ii),
we can eliminate a common invertible
factor from the $s$-th column in $\vert A\vert_{ij}$ where $s\neq j$.
(Note that this exponential function is actually invertible
as is shown in Eq. (\ref{inverse_exp}).)
Hence the $N$-soliton solution becomes the
following simple form where only the $I$-th column
is non-trivial, at $t\rightarrow +\infty$:
\begin{eqnarray*}
 \Phi_N \star f\rightarrow~~~~~~~~~~~~~~~~~~~~~~~~~~~~~~~~~~~~~~~~~~~~~~~~~~~~~~~~~~~~~~~~~~~~~~~~~~~~~~~~~~~~~~~~~~~~~~~~~~~~~~~~~\nonumber\\
\begin{array}{|cccccccc|}
1 & \!\cdots\!&1& e_\star^{\xi(\vec{x};k_I)}
+a_I e_\star^{-\xi(\vec{x};k_I)} &1&\!\cdots\!&1 & f\\
 -k_1&\! \cdots\!&-k_{I-1}
  &k_I(e_\star^{\xi(\vec{x};k_I)}
-a_I e_\star^{-\xi(\vec{x};k_I)})&k_{I+1}&\!\cdots\!&
k_N  &f^\prime\\
 \vdots&  &\vdots&\vdots &\vdots& & \vdots &\vdots\\
\! (-k_1)^{N-1}\!&\! \cdots\!&(-k_{I-1})^{N-1}\!&\! 
k_I^{N-1}(e_\star^{\xi(\vec{x};k_I)}
+(-1)^{N-1}
a_I e_\star^{-\xi(\vec{x};k_I)})\!
&k^{N-1}_{I+1}\!&\!\cdots\! &\! k_N^{N-1} \!&\! f^{(N-1)}\\
 (-k_1)^{N}& \!\cdots\!&(k_{I-1})^{N}& 
k_I^N(e_\star^{\xi(\vec{x};k_I)}
+(-1)^N a_I e_\star^{-\xi(\vec{x};k_I)})
 &k^{N}_{I+1}&\!\cdots\!&k_N^{N}& \fbox{$f^{(N)}$}\\
\end{array}~ ,
\end{eqnarray*}
and at $t\rightarrow -\infty$:
\begin{eqnarray*}
 \Phi_N \star f\rightarrow~~~~~~~~~~~~~~~~~~~~~~~~~~~~~~~~~~~~~~~~~~~~~~~~~~~~~~~~~~~~~~~~~~~~~~~~~~~~~~~~~~~~~~~~~~~~~~~~~~~~~~~~~\\
\begin{array}{|cccccccc|}
1 &\!\cdots\!&1& e_\star^{\xi(\vec{x};k_I)}
+a_I e_\star^{-\xi(\vec{x};k_I)} &1&\!\cdots\!&1 & f\\
 k_1&\!\cdots\!&k_{I-1}
  &k_I(e_\star^{\xi(\vec{x};k_I)}
-a_I e_\star^{-\xi(\vec{x};k_I)})&-k_{I+1}&\!\cdots\!&
-k_N  &f^\prime\\
 \vdots&  &\vdots&\vdots &\vdots& & \vdots &\vdots\\
 k_1^{N-1}\!&\!\cdots\!&\!k^{N-1}_{I-1}\!&\! 
k_I^{N-1}(e_\star^{\xi(\vec{x};k_I)}
+(-1)^{N-1}
a_I e_\star^{-\xi(\vec{x};k_I)})\!
&(-k_{I+1})^{N-1}\!&\!\cdots\! &\! (-k_N)^{N-1}\! &\!f^{(N-1)}\\
 k_1^{N}&\!\cdots\!&k^{N}_{I-1}& 
k_I^N(e_\star^{\xi(\vec{x};k_I)}
+(-1)^N a_I e_\star^{-\xi(\vec{x};k_I)})
 &(-k_{I+1})^{N}&\!\cdots\!&(-k_N)^{N}& \fbox{$f^{(N)}$}\\
\end{array}~ .
\end{eqnarray*}
Here we can see that all elements in between the first column and the
$N$-th column commute and depend only on $x + k_{I}^{2} t$
in $\xi(\vec{x};k_I)$, 
which implies that the corresponding asymptotic
configuration coincides with the commutative one, that is, 
the $I$-th one-soliton configuration with some
coordinate shift, so called the {phase shift}.
(We note that because $f$ is arbitrary,
there is no need to consider the products between
a column and the $(N+1)$-th column.
This observation for asymptotic behavior
can be made from Eq. (\ref{Nsol2}) also.) 
The commutative discussion has been studied in this way 
by many authors, and therefore, we conclude that
for the noncommutative KdV hierarchy,
{asymptotic behavior of the multi-soliton solutions 
is all the same as commutative one}, and as the results,
{the $N$-soliton solutions have $N$ isolated localized lump of energy 
and in the scattering process, they never decay and
preserve their shapes and velocities.
The phase shifts also appear by the same degree as commutative ones}.

\subsubsection{Asymptotic behavior of noncommutative KP solitons}

Next, let us focus on  
the asymptotic behavior of the $N$-soliton solutions to
the noncommutative KP equation 
where the space and time
coordinates are $(x,y,t)\equiv(x_1,x_2,x_{3})$
with the space-time noncommutativity $[x,t]_\star=i\theta$.
Here the constants $k_s$ and $k^{\prime}_s$
are non-zero real numbers and the constant $a_s$
will be redefined later. 

As we mentioned at the beginning of the present section,
one-soliton solutions are all the same as commutative ones.
However, we have to treat carefully for the noncommutative KP hierarchy.

First we comment on the the Baker-Campbell-Hausdorff (BCH) formula
for the the star exponential function $\xi(\xi;\vec{k})$ in the solution.
{}Let's focus on the noncommutative part of the 
star exponential function. The relevant part in the linear function
is $\xi(\xi;\vec{k})=k( x + k^2 t)$.
We note that we sometimes meet the following calculation:
\begin{eqnarray}
 e_\star^{\xi(\vec{x};k)}\star e_\star^{\xi(\vec{x};k^{\prime})}
=e^{(i/2)\theta(k k^{\prime 3}-k^3 k^{\prime})}
e_\star^{\xi(\vec{x};k)+\xi(\vec{x};k^{\prime})}
=e^{i\theta k k^{\prime}(k^{\prime 2}-k^2)}
e_\star^{\xi(\vec{x};k^{\prime})}\star e_\star^{\xi(\vec{x};k)}.
\label{BCH}
\end{eqnarray}
Let us see how we should eliminate the complex factor 
$\Delta:=(i/2)\theta k k^{\prime }(k^{\prime 2}-k^2)$
in the asymptotic region 
under the condition that the configurations take real values.

{}From Eq. (\ref{Nsol2}), naive one-soliton solution
can be expressed as follows: 
\begin{eqnarray}
 \label{1sol}
 u_2&=&\partial_x\left(\partial_x(e_\star^{\xi(\vec{x};k)}
+a e_\star^{\xi(\vec{x};k^{\prime})})\star (e_\star^{\xi(\vec{x};k)}
+a e_\star^{\xi(\vec{x};k^{\prime})})^{-1}\right)\nonumber\\
 &=&
  \partial_x\left((k^i+ak^{\prime i}
\Delta e_\star^{\eta(\vec{x};k,k^{\prime})})\star
(1+a \Delta
e_\star^{\eta(\vec{x};k,k^{\prime})})^{-1}\right),
\end{eqnarray}
where
 $\eta(\vec{x};k,k^{\prime}):=x(k^{\prime}-k)+y(k^{\prime 2}-k^2)
  +t(k^{\prime 3}-k^3)$. 
We note that the complex factor $\Delta$
cannot be absorbed by redefining a coordinate
such as $x\rightarrow x+(k^{\prime}-k)^{-1}\Delta$
because the space-time coordinates are real.\footnote{
In \cite{MaPa}, similar observations are made 
of the noncommutative Burgers equation where
the noncommutative parameter is not real but pure imaginary. 
This implies that the factor $\Delta$ is real 
and can be absorbed by a coordinate shift which affects the phase shift.} 
Instead of this, we redefine a positive real constant 
$\tilde{a}:=a \Delta$ in order to absorb the complex factor $\Delta$ 
so that $f_1=e_\star^{\xi(\vec{x};k)}
+a e_\star^{\xi(\vec{x};k^{\prime})}=\left(
1+\tilde{a} e_\star^{\eta(\vec{x};k,k^{\prime})}\right)
\star e_\star^{\xi(\vec{x};k)}$.
This avoids the coordinate shift by a complex number.
The configuration in asymptotic region is real. 

This point becomes important for scattering process of
the multi-soliton solutions.
We will soon see that the constants $a_s$ in the $N$-soliton solution
to the noncommutative KP equation 
should be replaced with a positive real number
$\tilde{a}_s$ which satisfies $a_s=\tilde{a}_s\Delta_s^{-1}$
where $\Delta_s:=e^{(i/2)\theta k_s k^{\prime}_s(k^{\prime 2}_s-k_s^2)}$.

Let us define new coordinates
comoving with the $I$-th soliton as follows:
\begin{eqnarray}
X:=x+k_I y+k_I^{2} t,~~~
Y:=x+k^{\prime}_I y+k^{\prime 2}_I t,
\label{comoving}
\end{eqnarray}
so that $X,Y$ are finite in the asymptotic region.
Then the function $\xi(x,y,t;k_s)$ can be rewritten
in terms of the new coordinates
as $\xi(X,Y,t;k_s)=A(k_s)X+B(k_s)Y+C(k_s)t$
where 
$A(k_s), B(k_s)$ and $C(k_s)$ are
real constants depending on $k_I,k^{\prime}_I$ and $k_s$.
We can get from Eq.
(\ref{comoving})
\begin{eqnarray}
 \left(
 \begin{array}{c}x\\y\end{array}
 \right)=\frac{1}{k^{\prime}_I-k_I}
  \left(
 \begin{array}{c}
 k^{\prime}_I X-k_I Y+k_Ik^{\prime}_I(k^{\prime}_I-k_I)t\\
 -X+Y+(k_I^2-k^{\prime 2}_I)t
 \end{array}
 \right),
\end{eqnarray}
and find
\begin{eqnarray*}
 \xi= x+k_sy+k_s^{n-1}t
  = \frac{k^{\prime}_I-k_s}{k^{\prime}_I-k_I}X
  +\frac{k_s-k_I}{k^{\prime}_I-k_I}Y
  +(k_s-k_I)(k_s-k^{\prime}_I)t.
\end{eqnarray*}
Here we assume that $C(k_s)\neq C(k^{\prime}_s)$
which corresponds to pure soliton scatterings.
(The condition $C(k_s)= C(k^{\prime}_s)$ could
lead to soliton resonances. For
commutative discussion, see e.g. \cite{OhWa}, and 
\cite{Kodama} as well.) 

Now let us take $t\rightarrow \pm \infty$ limit,
then, for the same reason as in the noncommutative KdV equation,
we can see that the asymptotic behavior of $f_s$ becomes:
\begin{eqnarray}
 f_s(\vec{x})\longrightarrow
\left\{\begin{array}{ll}
A_s e_\star^{\xi(\vec{x};\tilde{k}_{s})} &s\neq I\\
e_\star^{\xi(\vec{x};k_{I})}
+a_I e_\star^{\xi(\vec{x};k^{\prime}_{I})}&s=I
\end{array}
\right.
\end{eqnarray}
where $A_s$ is some real constant whose value is 1 or $a_s$,
and $\tilde{k}_s$ is a real constant
taking a value of $k_s$ or $k^{\prime}_s$.
As in the case of the noncommutative KdV equation,
the $s$-th ($s\neq I$) column
is proportional to a single exponential function
and we can eliminate this factor from the $s$-th column. 
Hence in the asymptotic region $t\rightarrow \pm \infty$,
the $N$-soliton solution becomes the
following simple form where only the $I$-th column
is non-trivial:
\begin{eqnarray*}
 \Phi_N \star f &\rightarrow& 
\begin{array}{|cccccccc|}
 1& \cdots &1& e_\star^{\xi(\vec{x};k_{I}) }
+a_I e_\star^{\xi(\vec{x};k^{\prime}_{I}) } &1&\cdots&1 & f\\
 \tilde{k}_1& \cdots &\tilde{k}_{I-1}&k_Ie_\star^{\xi(\vec{x};k_{I}) }
+a_Ik^{\prime}_I e_\star^{\xi(\vec{x};k^{\prime}_{I}) }
&\tilde{k}_{I+1}&
\cdots&
\tilde{k}_N  &f^\prime\\
 \vdots& & \vdots&\vdots &\vdots& & \vdots &\vdots\\
 \tilde{k}_1^{N-1}& \cdots &\tilde{k}^{N-1}_{I-1}& 
k_I^{N-1}e_\star^{\xi(\vec{x};k_{I}) }
+
a_I k^{\prime N-1}_I
e_\star^{\xi(\vec{x};k^{\prime}_{I})}  &\tilde{k}^{N-1}_{I+1}
&\cdots & \tilde{k}_N^{N-1} &f^{(N-1)}\\
 \tilde{k}_1^{N}& \cdots &\tilde{k}^{N}_{I-1}& 
k_I^Ne_\star^{\xi(\vec{x};k_I) }
+a_I k^{\prime N}_I e_\star^{\xi(\vec{x};k^{\prime}_{I}) }
 &\tilde{k}^{N}_{I+1} &\cdots& \tilde{k}_N^{N}& \fbox{$f^{(N)}$}\\
\end{array}~ \nonumber\\
&&=\begin{array}{|cccccccc|}
1& \cdots &1& 1
+\tilde{a}_I e_\star^{\eta(\vec{x};k_I,k^{\prime}_I) }&1 &\cdots&1& f\\
  \tilde{k}_1 & \cdots& \tilde{k}_{I-1}&k_I
+\tilde{a}_Ik^{\prime}_I e_\star^{\eta(\vec{x};k_I,k^{\prime}_{I}) } &\tilde{k}_{I+1}
& \cdots&
\tilde{k}_N  &f^\prime\\
 \vdots& & \vdots&\vdots &\vdots & & \vdots &\vdots\\
 \tilde{k}_1^{N-1}& \cdots &\tilde{k}^{N-1}_{I-1}& 
k_I^{N-1}
+
\tilde{a}_I k^{\prime N-1}_I
e_\star^{\eta(\vec{x};k_I,k^{\prime}_{I})}
 &\tilde{k}^{N+1}_{I+1}&\cdots & \tilde{k}_N^{N-1} &f^{(N-1)}\\
  \tilde{k}_1^{N}& \cdots &\tilde{k}^{N}_{I-1}& 
k_I^N
+\tilde{a}_I k^{\prime N}_I e_\star^{\eta(\vec{x}; k_I,k^{\prime}_{I}) }
  &\tilde{k}^{N}_{I+1}&\cdots& \tilde{k}_N^{N}& \fbox{$f^{(N)}$}\\
\end{array}~.
\end{eqnarray*}
Here we can see that all elements between the first column and the
$N$-th column are real and depend only on
$x(k^{\prime }_I-k_{I})+t(k^{\prime 3}_I-k_{I}^{3})$
for noncommutative coordinates.
This implies that the corresponding asymptotic
configuration coincides with the commutative one.
Hence, we can also conclude that for the noncommutative KP equation,
{asymptotic behavior of the multi-soliton solutions 
is all the same as commutative one in the process of
pure soliton scatterings}.
As the results,
{the $N$-soliton solutions possess $N$ isolated localized lump of energy 
and in the pure scattering process, they never decay and
preserve their shapes and velocities of the localized solitary waves.}
This coincides with the result on 2-soliton scattering 
studied by Paniak \cite{Paniak}.

\bigskip
As is suggested by  
the stability of the $N$-soliton solution, 
there actually exist infinite conserved densities
of the noncommutative  KP equation with space-time noncommutativity:
\begin{eqnarray}
  \sigma_n
={\mbox{coef}}_{-1} L^n
-3\theta
\left(({\mbox{coef}}_{-1}L^n)\diamond u_3^\prime
+({\mbox{coef}}_{-2}L^n)\diamond u_2^\prime
\right),~~~~n=1,2,\cdots 
\end{eqnarray}
where coef$_{-l} L^n$ denotes the coefficient of $\partial^{-l}$
in $L^n$. (In particular, coef$_{-1} L^n$ is the residue of $L^n$.)
The product ``$\diamond$'' is called the {Strachan's product}
\cite{Strachan} and defined by
\begin{eqnarray}
 f(x)\diamond g(x)
:=\sum_{p=0}^{\infty}
\fr{(-1)^p}{(2p+1)!}\left(\frac{1}{2}\theta^{ij}
\del_i^{(x^\prime)}\del_j^{(x^{\prime\prime})}\right)^{2p}
f(x^\prime)g(x^{\prime\prime})\vvert_{x^\prime=x^{\prime\prime}=x}.
\end{eqnarray}
This is a commutative and non-associative product.
Conserved densities
for one-soliton configuration are not deformed
in the noncommutative extension because one soliton solutions
can be always reduced to commutative ones

\section{NC ASDYM Hierarchy and Soliton Solutions}

Finally, we present noncommutative anti-self-dual Yang-Mills (ASDYM)
equations in 4 dimension and its hierarchy generalization.
Let $(x_0,x_1,x_2,\cdots; y_0, y_1, y_2,\cdots)$
be complex coordinates and define covariant derivatives
$D_{x_k}:=\partial_{x_k} + A_{x_k} , D_{y_l}:=\partial_{y_l} + A_{y_l}
~(k,l=0,1,2,\cdots)$
where $A_{x_k}$ and $A_{y_l}$ are $n\times n$ complex matrices.
Noncommutativity is introduced into the coordinates.

Let us consider the following linear systems:
\begin{eqnarray}
 L_k \star \psi &:=& (D_{x_k}-\zeta D_{x_{k-1}})\star \psi,\\
 M_l \star \psi &:=& (D_{y_l}-\zeta D_{y_{l-1}})\star \psi,
\end{eqnarray}
where $\zeta\in {\mathbb{C}} P^1$ is a spectral parameter
which commutes with all spatial coordinates.
The compatibility condition of the linear systems is
$[L_k,M_l]_\star=0$ for any $k,l$.
This yields an infinite systems of partial differential equations, 
which is called the noncommutative anti-self-dual Yang-Mills hierarchy.
By identification of $x_0\equiv \bar{z}, x_1\equiv w,
y_0\equiv -\bar{w}, y_1\equiv z$, the compatibility
condition for $k=l=1$ coincides with the
noncommutative anti-self-dual Yang-Mills equation \eqref{ncASDYM}.
For the commutative anti-self-dual Yang-Mills hierarchies, 
see e.g. \cite{ACT, MaWo, Nakamura, Suzuki, Takasaki_CMP}.

The noncommutative anti-self-dual Yang-Mills
hierarchy equations can be rewritten as
the following form:
\begin{eqnarray}
 \label{yang}
  \partial_{y_{l}} (J^{-1} \star \partial_{x_{k-1}} J)
  -\partial_{x_{k}} (J^{-1} \star \partial_{y_{l-1}} J)=0,
\end{eqnarray}
where $J$ is an $n\times n$ matrix. 
The equation \eqref{yang} and the matrix $J$
is called the noncommutative Yang's hierarchy equation and 
the Yang's $J$-matrix, respectively.
For $k=l=1$, this coincides with the noncommutative Yang's equation.

Anti-self-dual gauge fields can be reproduced from the
solution $J$ to the equation \eqref{yang} by decomposition 
$J=\tilde{h}^{-1} \star h$ as 
\begin{eqnarray*}
\label{a}
A_{y_l}\!=-(\partial_{y_l} h)\star h^{-1}\!,~A_{x_k}\!=-(\partial_{x_k} h)\star h^{-1}\!,~
A_{x_{k-1}}\!=-(\partial_{x_{k-1}}\widetilde{h})\star \widetilde{h}^{-1}\!,~
A_{y_{l-1}}\!=-(\partial_{y_{l-1}}\widetilde{h})\star \widetilde{h}^{-1}.
\end{eqnarray*}
The proof is the same as the noncommutative anti-self-dual Yang-Mills
equation. (See e.g. \cite{GHN}.)

{}From now on, we focus on the $n=2$ case.
The $J$-matrix can be reparametrized without loss of generality
as follows:
\begin{eqnarray}
 \label{J}
 J=\left[\begin{array}{cc} p -r \star q^{-1} \star s&-r \star q^{-1}
   \\ q^{-1}\star   s &q^{-1}\end{array}
   \right].
\end{eqnarray}
Exact solutions of the $m$-th Atiyah-Ward ansatz
are represented in terms of quasideterminants ($m=0,1,2,\cdots$):
\begin{eqnarray}
\label{AWsol}
p_m&=&
\begin{array}{|cccc|}
\fbox{$\varphi_0$}&\varphi_{-1} & \cdots & \varphi_{-m}\\
\varphi_1 &\varphi_0&\cdots & \varphi_{1-m} \\
\vdots &\vdots &\ddots & \vdots\\
\varphi_{m} &\varphi_{m-1} &\cdots &\varphi_0
\end{array}^{-1},~
q_m=
\begin{array}{|cccc|}
\varphi_0&\varphi_{-1} & \cdots & \varphi_{-m}\\
\varphi_1 &\varphi_0&\cdots & \varphi_{1-m} \\
\vdots &\vdots &\ddots & \vdots\\
\varphi_{m} &\varphi_{m-1} &\cdots &\fbox{$\varphi_0$} 
\end{array}^{-1},\nonumber\\
r_m&=&
\begin{array}{|cccc|}
\varphi_0&\varphi_{-1} & \cdots & \varphi_{-m}\\
\varphi_1 &\varphi_0&\cdots & \varphi_{1-m} \\
\vdots &\vdots &\ddots & \vdots\\
\fbox{$\varphi_{m}$} &\varphi_{m-1} &\cdots &\varphi_0 
\end{array}^{-1},~
s_m=
\begin{array}{|cccc|}
\varphi_0&\varphi_{-1} & \cdots & \fbox{$\varphi_{-m}$}\\
\varphi_1 &\varphi_0&\cdots & \varphi_{1-m} \\
\vdots &\vdots &\ddots & \vdots\\
\varphi_{m} &\varphi_{m-1} &\cdots &\varphi_0
\end{array}^{-1}.
\end{eqnarray}
where the scalar functions $\varphi_i(x;y)$ can be 
determined from a scalar function $\varphi_0$
recursively by the chasing relation:
\begin{eqnarray}
\label{chasing}
 \frac{\partial \varphi_i}{\partial {y_l}}
= -\frac{\partial \varphi_{i+1}}{\partial {y_{l-1}}},~~~
 \frac{\partial \varphi_i}{\partial {x_k}}
= -\frac{\partial \varphi_{i+1}}{\partial x_{k-1}},~~~~~~~
-m\leq i\leq m-1~~~(m\geq 2).
\end{eqnarray}
The scalar function $\varphi_0$ is a solution to
the linear equation $(\partial x_{k} \partial y_{k-1}
-\partial y_{k} \partial x_{k-1})\varphi_0=0$.

\subsection{Asymptotic behavior of noncommutative ASDYM solitons}

Here let us discuss 
$N$ soliton solutions to the anti-self-dual Yang-Mills hierarchy \eqref{yang}
and asymptotic behaviors of the 2-soliton case.
In order to discuss it, 
we have to pick a specified coordinate up 
in order to identify it with time coordinate. 

By the identification of $x_{k-1}\equiv \bar{z}=t_{1}-it_{2},
x_{k}\equiv w=t_{3}+it_{4},
y_{l-1}\equiv -\bar{w}=-t_{3}-it_{4}, y_{l}\equiv z=t_{1}+it_{2}$
where $t_\mu~(\mu=1,2,3,4)$ is a real coordinate,
the linear equation becomes the Laplace equation 
in 4-dimension:
\begin{eqnarray}
 \label{laplace}
\partial^2 \varphi_0(t)=0,
\end{eqnarray}
where $\partial^2:=\partial_\mu \partial^\mu=\partial_1^2 +
\partial_2^2+\partial_3^2+\partial_4^2$.
The Yang's equation \eqref{yang} becomes 
\begin{eqnarray}
\label{yang_eq}
 \partial_z(J^{-1} \star \partial_{\bar{z}} J)
 +\partial_w (J^{-1} \star \partial_{\bar{w}} J)=0.
\end{eqnarray}

The following solution to the Laplace equation \eqref{laplace}
leads to an $N$ soliton solution to the anti-self-dual Yang-Mills
hierarchy equation: 
\begin{eqnarray*}
\label{Nsoliton}
\varphi_0(t)
  &=&1+\sum_{s=1}^{N} a_s e^{\xi(k_{(s)};t)}
+\sum_{i_1=1,i_2=1, i_1<i_2}^{N}
a_{i_1i_2} 
e^{\xi(k_{(i_1)};t)+\xi(k_{(i_2)};t)}
+
\cdots\\
&&+
a_{1 2 \cdots N} 
e^{\xi(k_{(1)};t)+\xi(k_{(2)};t)+\cdots+\xi(k_{(N)};t)}
,~~~\xi(k;t):=k_{\mu} t^\mu
\end{eqnarray*}
where the coefficients $a_{i_1\cdots i_s}~(s=1,\cdots,N)$
are complex constants
and $k_{(s)\mu}~(\mu=1,2,3,4)$ are real parameters 
which satisfy $k_{(s)\mu} k_{(s)}^\mu=0$.
The commutative limit of this $N$-soliton solution
reduces to the $N$ non-linear plane wave solution \cite{deVega}. 
We note that other scalar functions 
can be the same representation as \eqref{Nsoliton}
because of the chasing relation \eqref{chasing}.

We note that the coefficients 
$a_{i_1\cdots i_s}~(s=1,\cdots,N)$
are in general not real but complex,
because there is a gauge freedom:
$A_\mu \mapsto g^{-1}\star A_\mu
\star g +g^{-1}\star\partial_\mu  g$. 
Here we focus, however, on real-valued
configurations in order to compare with
the previous discussion. 
We finally need to check the asymptotic behavior of
gauge invariant quantities such as 
 $\displaystyle \int d^4 t \, {\mbox{Tr}}\, F^\star_{\mu\nu}$ and $\displaystyle
 \int d^4 t \, {\mbox{Tr}}\, F^\star_{\mu\nu} \star F^\star_{\mu\nu}$

Let us discuss the asymptotic behavior of the 
$N$ soliton solutions of the anti-self-dual Yang-Mills equation, 
which is called the ASDYM solitons. 
Now noncommutativity is assumed to be introduced into
a spatial coordinate $x\equiv t_1$ and time coordinate $t\equiv t_3$
such that $[x,t]_\star= i\theta$.
We consider the $t\rightarrow \pm \infty$ limit.

One soliton solution is given by
\begin{eqnarray}
\varphi_0  =1+ a e^{\xi(k;t)}.
\end{eqnarray}
Dependence of noncommutative coordinates is 
$\varphi_0(x+vt)~(v:=k_3/k_1)$ and hence 
the configuration reduces to the commutative one.
This can be interpreted as a domain wall in 4-dimension. 
D-brane interpretation of this solution is worth studying.

Two soliton solution is given by 
\begin{eqnarray}
\label{2soliton}
\varphi_0
 =1
 + a_1 e^{\xi(k;t)}
 + a_2 e^{\xi(k^\prime;t)}
 + a_{12} e^{\xi(k;t)+\xi(k^\prime;t)}.
\end{eqnarray}
Let us ride on the comoving frame with the first 
soliton so that $k_\mu t^\mu$ and $e^{\xi(k;t)}$ are finite.
In the limit of $t\rightarrow \pm \infty$, 
the term $e^{\xi(k^\prime;t)}$ goes to 0 or infinity. 
Hence in the asymptotic region, 
the 2-soliton solution becomes the following case (i) or (ii):
\begin{eqnarray*}
\label{varphi_0}
\varphi_0 
\longrightarrow 
\left\{\begin{array}{ll}
{\mbox{(i) }}  &1 + a_1 e^{\xi(k;t)} \\
{\mbox{(ii) }}  & a_2 e^{\xi(k^\prime;t)}+a_{12}  e^{\xi(k;t)+\xi(k^\prime;t)} = 
(a_2+ \tilde{a}_{12} e^{\xi(k;t)})\star e^{\xi(k^\prime;t)}
\end{array}
\right.
\end{eqnarray*}
where $a_{12}=\tilde{a}_{12}\Delta$
and $\Delta:=e^{(i/2)\theta (k_1 k_3^\prime-k_3 k_1^\prime)}$.
We assume that $\tilde{a}_{12}$ is real. 

In the case of (i), we can find that 
$\varphi_0(t,x)=\varphi_0(x+vt)$ 
and hence the configuration coincides with commutative one.
In the case of (ii), we have to proceed the calculation. 
As is commented, other scalar functions in 
the Atiyah-Ward ansatz solution \eqref{AWsol} have the form: 
$\varphi_i
 = b_i
 + c_i e^{\xi(k;t)}
 + d_i e^{\xi(k^\prime;t)}
 + r_i e^{\xi(k;t)+\xi(k^\prime;t)}$ 
where $b_i, c_i, d_i, r_i$ are constants,
and $b_i, c_i, d_i$ are real.
Asymptotic behavior of (ii) is 
$\varphi_i 
\longrightarrow  
(c_i+ \tilde{r}_{i} e^{\xi(k;t)})\star e^{\xi(k^\prime;t)}$.
where $r_{i}=\tilde{r}_{i}\Delta$
so that $\tilde{r}_{i}$ is real. 

Because of the multiplication property of columns
of quasideterminants, the Atiyah-Ward ansatz solution \eqref{AWsol}
have the common asymptotic form:
$p,q,r,s \rightarrow f(t+vx) \star e^{\xi(k^\prime;t)}$.

The gauge fields can be recovered 
from the matrices $h$ and $\widetilde{h}$ as in \eqref{a}.
Let us decompose the matrix $J$ into $h$ and $\tilde{h}$ as follows:
\begin{eqnarray*}
J =
\left[\begin{array}{cc} p\star  -r\star  q^{-1} s&-r\star q^{-1}
   \\ q^{-1} \star s &q^{-1}\end{array}
   \right]
=\left[\begin{array}{cc}1&r
   \\ 0 &q\end{array}
   \right]^{-1}\star
\left[\begin{array}{cc}
	     p&0
\\ s&1\end{array}
   \right]=\widetilde{h}^{-1}\star  h.
\label{M-W}
\end{eqnarray*}
The gauge fields are calculated as 
\begin{eqnarray*}
\label{a}
 &&A_{z}\!=\!
-(\partial_z h)\star h^{-1}\!=\!
\left[\begin{array}{cc}
	     -(\partial_z p)\star p^{-1}&0
\\ -(\partial_z s)\star p^{-1}&0\end{array}
   \right], ~  
A_{w}\!=\!-(\partial_w h)\star h^{-1}\!=\!
\left[\begin{array}{cc}
	     -(\partial_w p)\star p^{-1}&0
\\ -(\partial_w s)\star p^{-1}&0\end{array}
   \right],\nonumber\\
&&
A_{\widetilde{z}}\!=\!-(\partial_{\widetilde{z}}\widetilde{h})\star\widetilde{h}^{-1}
\!=\!
\left[\begin{array}{cc}
0&	     -(\partial_{\widetilde{z}} r)\star q^{-1}
\\
0& -(\partial_{\widetilde{z}} q)\star q^{-1}\end{array}
   \right],~
A_{\widetilde{w}}\!=\!-(\partial_{\widetilde{w}}\widetilde{h})\star\widetilde{h}^{-1}
\!=\!
\left[\begin{array}{cc}
0&	     -(\partial_{\widetilde{w}} r)\star q^{-1}
\\
0& -(\partial_{\widetilde{w}} q)\star q^{-1}\end{array}
   \right].
\end{eqnarray*}
We can see that the common factor $e^{\xi(k^\prime;t)}$ in
$p,q,r,s$ is canceled out here, and the coordinate dependence in the 
gauge fields becomes $A_{\mu}(t,x)=A_{\mu}(x+v t)$.
Note that there is no difference between commutative case
and noncommutative case in the derivation from \eqref{varphi_0}.
We can therefore conclude the gauge invariant quantities
consist of $F^\star_{\mu\nu}$
are the same as commutative ones.

Let us consider the comoving frame with the second
soliton where $k^\prime_\mu t^\mu$ and $e^{\xi(k^\prime;t)}$ are finite.
In the limit of $t\rightarrow \pm \infty$, 
the factor $e^{\xi(k;t)}$ goes to (i) 0 or (ii) infinity.
The case (i) reduces to one-soliton configuration.
The case (ii) leads to, in similar way, 
the following asymptotic behaviors of the scalar functions:
$\varphi_i 
\rightarrow  
e^{\xi(k;t)}\star (a_1+ \tilde{a}_{12} e^{\xi(k^\prime;t)})$,
$\varphi_i 
\rightarrow  
e^{\xi(k;t)}\star (d_i+ \tilde{r}_{i} e^{\xi(k^\prime;t)})$,
and $p,q,r,s \rightarrow e^{\xi(k;t)} \star f(x+v^\prime t) $.
We note that the coefficients $a_1, \tilde{a}_{12}, d_i, \tilde{r}_i$ 
are the same as those in case (i).) We can see that the common factor
$e^{\xi(k;t)}$ appears in the gauge fields as
$A_\mu(x,t) \rightarrow e^{\xi(k;t)} \star A(x+v^\prime t) \star e^{-\xi(k;t)}$
This is essentially gauge equivalent to $A(x+v^\prime t)$
up to constants which do not contribute the
field strength. 
Hence 
the gauge invariant quantities consist of $F^\star_{\mu\nu}$
are the same as commutative ones in this case as well.

Therefore we can conclude that 
the asymptotic behavior of the 2-soliton solutions 
is the same as commutative one \cite{deVega}
and as the results, 
the $2$-soliton solutions has $2$ isolated localized lump of energy
and in the scattering process, they never decay and
preserve their shapes and velocities of the localized solitary waves.

Higher-charge soliton scattering is worth studying.
For this purpose, Wronskian-type solutions \cite{GNO}
would be suitable. This will be reported elsewhere.


\subsection*{Acknowledgments}

One of the authors would like to thank
the organizers at the workshop on 
Physics and Mathematics of Nonlinear Phenomena
(PMNP2017): 50 years of IST in Gallipoli, Italy.
The work of MH was supported 
by Grant-in-Aid for Scientific Research (\#16K05318).

\end{document}